\documentstyle[12pt,amssymb,epsfig]{article}
\begin{document}
%%%%%%%%%%%%%%%%
\setlength{\baselineskip}{0.65cm}
\setlength{\parskip}{0.35cm}
\def    \beq             {\begin{equation}}
\def    \eeq             {\end{equation}}
\def    \beqa            {\begin{eqnarray}}
\def    \eeqa            {\end{eqnarray}}
\def \e {{\rm e}}
\def \as {{\alpha_s}}
%%%%%%%%%%%%%%%%%
\begin{titlepage}
%%%%%%%%%%%%%%%%%
%
\begin{flushright}
BNL-NT-01/14 \\
RBRC-193 \\
July 2001
\end{flushright}

\vspace*{1.5cm}
\begin{center}
\LARGE
{\bf {Towards a global analysis}} 

\medskip
{\bf {of polarized parton distributions}}

\vspace*{2cm}
\large 
{Marco Stratmann$^{a,b}$ and Werner Vogelsang$^{c,d}$}

\vspace*{1.0cm}
\normalsize
{\em $^a$Institut f\"ur Theoretische Physik, Universit\"at Regensburg,\\
D-93040 Regensburg, Germany}\\

{\em $^b$C.N.\ Yang Institute for Theoretical Physics, SUNY at Stony Brook,\\
Stony Brook, New York 11794 -- 3840, U.S.A.}

\vspace*{0.5cm}
{\em $^c$Physics Department, Brookhaven National Laboratory,\\
Upton, New York 11973, U.S.A.}\\

{\em $^d$RIKEN-BNL Research Center, Bldg. 510a, Brookhaven National Laboratory, \\
Upton, New York 11973 -- 5000, U.S.A.}
\end{center}

\vspace*{1.5cm}
\begin{abstract}
\noindent

We present a technique for implementing in a fast 
way, and without any approximations, higher-order calculations 
of partonic cross sections into global analyses of parton distribution 
functions. The approach, which is set up in Mellin-moment space, is 
particularly suited for analyses of future data from polarized 
proton-proton collisions, but not limited to this case.
The usefulness and practicability of this method is demonstrated
for the semi-inclusive production of hadrons in deep-inelastic
scattering and the transverse momentum distribution of
``prompt'' photons in $pp$ collisions,
and a case study for a future global analysis of polarized parton
densities is presented.
\end{abstract}
\end{titlepage}
\newpage
%
%%%%%%%%%%%%%%%%%%%%%%
\section{Introduction}
%%%%%%%%%%%%%%%%%%%%%%
%
\noindent
High-energy spin physics has been going through a period of great popularity
and rapid developments ever since the measurement of the proton's 
spin-dependent deep-inelastic structure function $g_1^p$ by the 
EMC~\cite{emc} more than a decade ago. As a result of combined experimental 
and theoretical efforts, we have gained some fairly precise information 
concerning, for example, the total quark spin contribution to the nucleon
spin. Yet, many other interesting and important questions, most of which 
came up in the wake of the EMC measurement, remain unanswered so far, the 
most prominent ``unknown'' being the nucleon's spin-dependent gluon density,
$\Delta g$. Also, polarized {\em inclusive} deep-inelastic scattering (DIS) data
do not provide enough information for a complete separation of the
distributions for the
different quark and anti-quark flavors $u,\bar{u},d,\bar{d},s$, and $\bar{s}$. 
Here in particular a possible flavor asymmetry in the nucleon's light sea, 
$\Delta \bar{u} - \Delta \bar{d} \neq 0$, has attracted quite some
interest, and several models have been proposed
recently \cite{ref:su2br1,ref:su2br2,ref:su2br3}.
Current and future dedicated spin experiments are expected to 
vastly broaden our understanding of the nucleon spin structure by
studying reactions that give further access to its spin-dependent
parton distributions, among them $\Delta g$ and $\Delta \bar{u}$, $\Delta \bar{d}$.
In addition to lepton-nucleon 
scattering, there will also be for the first time information coming
from very inelastic polarized $pp$ collisions at the BNL Relativistic 
Heavy-Ion Collider RHIC~\cite{rhic}.

Having available at some point in the near future spin data on 
various different reactions, one needs to tackle the question 
of how to determine the polarized parton densities from the data.
Of course, this problem is not at all new: in the unpolarized
case, several groups perform such ``global analyses'' of the
plethora of data available there~\cite{cteq,grv}. 
The strategy is in principle clear: an ansatz for the parton distributions 
at some initial scale $\mu_0$, given in terms of appropriate functional 
forms with a set of free parameters, is evolved to a scale 
$\mu_F$ relevant for a certain data point for a certain cross section. 
Then the parton densities at scale $\mu_F$ are used to compute the 
theoretical prediction for the cross section, 
and a $\chi^2$ value is assigned that represents the quality of the 
comparison to the experimental point. This is done for all data 
points to be included in the analysis, and subsequently the parameters 
in the ansatz for the parton distribution functions are varied, until 
eventually a minimum in $\chi^2$ is reached. 

In practice, this approach is not fully viable if the partonic scattering 
is treated beyond the lowest order of perturbation theory.  
The numerical evaluation of the hadronic cross section at higher 
orders is usually a rather time-consuming procedure as it often requires 
several tedious numerical integrations, not only for the 
convolutions with the parton densities, 
but also for the phase space integrations in the partonic
cross section. The fitting procedure outlined above, on the other
hand, usually requires thousands of computations of the cross 
section for any given data point, and so the computing 
time required for a fit easily becomes excessive even on modern
workstations. We note that for practically all 
reactions of interest in the unpolarized and polarized cases
the first-order QCD corrections to the respective partonic 
cross sections are known by now. They are generally
indispensable in order to arrive at a firmer theoretical 
prediction for hadronic cross sections; for instance the
dependence on the unphysical factorization and renormalization 
scales is reduced when going to higher orders in the perturbative
expansion. Only then can one reliably extract information on the 
parton distribution functions.

In the unpolarized case, a way to get around this problem is 
based on the fact that the parton densities are already known 
here rather accurately~\cite{cteq,grv}. Their gross features
are basically determined by the wealth of very precise DIS data 
which cover a wide kinematical range in the momentum fraction
$x$ and the scale $\mu_F\simeq Q$.
As a consequence, the theory answer for a certain
cross section is expected to change in a very predictable way 
when going from the lowest-order Born level to the 
first-order approximation.
It is then possible to pre-calculate a set of correction factors $K_i$
($i$ running over the data points), 
and to simply multiply them in each step of the fitting procedure to
the {\em lowest-order} approximation for the cross section, the latter 
being usually much faster to evaluate than that involving 
higher order terms. The $K_i$ usually hardly change at all from 
one set of parton distributions to another, and in any case one 
may update them if necessary at certain stages of the fitting
procedure. 

It should be noted, however, that this way of treating 
next-to-leading order (NLO) corrections in a fit
is not necessarily adequate in all cases of interest. In
particular if one is interested in extracting information
about the gluon density at large values of $x$, where it is
only rather poorly constrained at the moment,
the correction factors  $K_i$ cannot be reliably pre-calculated,
and they may vary considerably during the fitting procedure.
It is therefore desirable to incorporate NLO cross sections
without any approximations in future analyses of parton densities.

In the polarized case it is in general not at all clear 
whether a strategy based on correction factors $K_i$ will work. 
Here, the parton densities are known 
with {\em much} less accuracy so far.  
It is therefore not possible to pre-calculate higher-order correction 
factors that one would be able to keep fixed throughout the fit, while using
``fast'' lowest-order expressions for the partonic cross sections. 
For instance, even though it is well known that for a sizable 
$\Delta g$ the $q+g\to \gamma +q$ Compton subprocess is the dominant 
contributor to the transverse momentum distribution
of a ``prompt'' photon in the kinematical 
region of interest, this is by no means the case if $\Delta g$ 
happens to be small, in which case all other channels, even 
genuine NLO ones, may become equally important. 
In addition, the spin-dependent parton distributions, 
as well as the polarized partonic cross sections, may have zeros in the 
kinematical regions of interest, near which the predictions at
lowest order and the next order will show marked differences.
Therefore, even if the correction factors are updated at 
times during the fitting procedure, the convergence of the fit is not 
warranted. Conversely, if one updates the $K_i$ frequently, the fit 
will become too slow again.

Clearly, in the polarized case, the goal {\em must} be to find a way
of implementing efficiently, and without approximations, the
{\em exact} NLO expression for any hadronic cross section such as 
the prompt photon cross section into the fitting procedure. 
As will be shown in the next Section, this can be achieved
in a very simple and straightforward way by going to Mellin-$n$
moment space. A technique of this sort was first used for the case
of jet production in DIS as a means of extracting information 
about the unpolarized gluon density \cite{bghv}.
The relevant generalization to hadron-hadron scattering, which 
is more involved and requires a ``double Mellin transform''  
was recently provided in \cite{ref:kosower}. However,
\cite{ref:kosower} focuses on the formalism and the technical
aspects of the Mellin transformation, rather than on
its actual practicability, and the usefulness in a global QCD 
analysis has never been demonstrated. 

Before we demonstrate in some detail the potential of the Mellin technique
in praxis for two examples
relevant for future global analyses of polarized parton densities, 
which is the main thrust of this paper,
we start off in the next Section by 
rederiving the required formalism in an easy and transparent way. 
In Section 3 we will consider first the semi-inclusive production of hadrons in
polarized DIS as the simplest application of the
Mellin technique. The $n$ moments of the partonic
cross sections can be taken analytically in this case.
Due to the subsequent fragmentation of a final state parton 
into the observed hadron, semi-inclusive DIS (SIDIS) is sensitive
to different flavor combinations than inclusive DIS data.
It has also the advantage that we have already
data at our disposal \cite{ref:smc,ref:hermes}
which can be analysed in 
terms of a possible flavor asymmetry $\Delta \bar{u} - \Delta \bar{d}$
of the light sea.
As a second example we study the production
of a prompt photon at high transverse momentum $p_T$ in
$pp$ collisions at RHIC in Section 4. Its sensitivity
to the gluon distribution via the LO Compton subprocess, which, 
along with the cleanliness of the prompt
photon signal, is the reason why this process will be the flagship
measurement of $\Delta g$ at RHIC~\cite{rhic}.
As a first case study for future global analyses 
we also carry out a ``toy'' analysis of DIS {\em and}
projected prompt photon data to highlight the power of future
RHIC $pp$ data to pin down $\Delta g$.
We briefly summarize the main results in Section 5.

%%%%%%%%%%%%%%%%%%%%%%%%%%%%%%%%%%%%%%%%%%%%%%%%%%%%%%%%%%%%%%%%%
\section{Hadronic Cross Sections and the Mellin Moment Technique}
%%%%%%%%%%%%%%%%%%%%%%%%%%%%%%%%%%%%%%%%%%%%%%%%%%%%%%%%%%%%%%%%%
\noindent
The factorization theorem~\cite{fact} ensures that in the presence of a 
hard scale in a reaction the corresponding (spin-dependent) 
hadronic cross section 
can be written as a sum over ``convolutions'' of parton 
densities with partonic hard-scattering cross sections. The latter
are perturbatively calculable and are specific to the 
reaction under consideration. The parton distributions, which for
spin-dependent interactions contain the desired information on
the nucleon's spin structure, depend on long-distance phenomena.
However, they are universal: a single set of distributions 
for (anti-)quarks $u,\bar{u},d,\bar{d},s,\bar{s},\ldots$ and gluons $g$,  
predicts all data sets simultaneously. 

To be specific, for a general spin-dependent cross section
in longitudinally polarized $pp$ collisions,
differential in a certain observable $O$ and integrated over experimental 
bins in other kinematical variables $T$, one has
\begin{eqnarray} 
\label{eq1}
\frac{d\Delta \sigma^H}{dO} &\equiv&
\frac{1}{2} \left[\frac{d\sigma^H}{dO}(++) - 
\frac{d\sigma^H}{dO}(+-)
\right] \\[2mm]
&=& 
\sum_{a,b,c}\, \int_{\mathrm exp-bin} dT\, 
\int_{x_a^{\mathrm min}}^1 dx_a 
\int_{x_b^{\mathrm min}}^1 dx_b 
\int_{z_c^{\mathrm min}}^1 dz_c \,\,
\Delta f_a (x_a,\mu_F) \,\Delta f_b (x_b,\mu_F) 
D_c^H(z_c,\mu_F') \, \nonumber \\ [2mm]
&&
\times \,\frac{d\Delta \hat{\sigma}_{ab}^{c}}{dO dT}
(x_aP_A, x_bP_B, P_H/z_c, T, \mu_R, \mu_F, \mu_F') \; ,\nonumber
\end{eqnarray}
where the arguments $(++)$ and $(+-)$ in the first line
of Eq.~(\ref{eq1}) refer to the helicities of the incoming hadrons
$A$ and $B$. The
$\Delta f_i$ are the spin-dependent parton distributions,
defined as
\begin{equation}\label{eq2}
\Delta f_i(x,\mu_F) \equiv f_i^+(x,\mu_F) -  f_i^-(x,\mu_F) \; ,
\end{equation}
where $f_i^+$ ($f_i^-$) denotes the number density of a parton-type 
$f_i$ with helicity `+' (`$-$') in a proton with positive helicity,
carrying the fraction $x$ of the proton's momentum. 
The $D_c^H(z,\mu_F')$ represent the unpolarized fragmentation functions.
They parameterize the probability that a parton $c$ fragments into
the observed final state $H$, e.g., a charged pion, with momentum $P_H=z\,p_c$.
For some observables, such as (di-)jets,
there is no need for a fragmentation function in Eq.~(\ref{eq1}).

The scales $\mu_F$ and $\mu_F'$ are the factorization scales for
initial and final state collinear singularities, respectively,
and reflect the certain amount of arbitrariness in the separation 
of short-distance and long-distance physics embodied in Eq.~(\ref{eq1}).
Even though the parton densities (fragmentation functions)
cannot presently be derived from first principles,
their dependence on $\mu_F$ ($\mu_F'$) is calculable perturbatively 
in terms of the ``DGLAP'' evolution equations~\cite{dglap}, allowing to relate
their values at one scale to their values at any other $\mu_F$
($\mu_F'$). The other scale, $\mu_R$, in Eq.~(\ref{eq1})
is the renormalization scale, introduced in the procedure of 
renormalizing the strong coupling constant. Finally, 
the sum in Eq.~(\ref{eq1}) is over all 
contributing partonic channels $a+b\to
c + X$, with 
$d\Delta \hat{\sigma}_{ab}^{c}$ the associated partonic cross
section, defined in complete analogy with the first line of 
Eq.~(\ref{eq1}), the helicities now referring to partonic ones: 
\begin{equation} \label{eq3}
d \Delta\hat{\sigma}_{ab}^{c} \equiv
\frac{1}{2} \Bigg[ d\hat{\sigma}_{ab}^{c}(++) - 
d\hat{\sigma}_{ab}^{c}(+-)
\Bigg] \; .
\end{equation}
As mentioned earlier, the $d\Delta \hat{\sigma}_{ab}^{c}$
are perturbative, that is, they have the expansion
\begin{equation} \label{eq4}
d\Delta \hat{\sigma}_{ab}^{c} = 
d\Delta \hat{\sigma}_{ab}^{c,(0)} + 
\left( \frac{\alpha_s}{\pi} \right)
d\Delta \hat{\sigma}_{ab}^{c,(1)} + 
\left( \frac{\alpha_s}{\pi} \right)^2
d\Delta \hat{\sigma}_{ab}^{c,(2)} + \ldots \;\,.
\end{equation}

It should be noted that lepton-hadron reactions are also 
included in Eq.~(\ref{eq1}) by simply setting
$\Delta f_b(x_b,\mu_F)=\delta(1-x_b)$. We will consider this example
in some detail in Section 3 as it is the simplest application of the
Mellin moment technique which we are going to advocate in the
following as a straightforward tool to extract information about parton
densities from a global QCD analysis.

For the polarized parton distribution functions,
the Mellin moments are defined as
\beq
\Delta f_i^n(\mu) \equiv \int_0^1 dx \;x^{n-1}  \Delta f_i(x,\mu) \; .
\eeq
It is well known~\cite{grv90} that the evolution equations for the 
the parton densities become particularly simple in Mellin-$n$
space, since the convolutions occuring in the $x$-space 
equations factorize into simple products under moments. This 
allows for a straightforward analytic solution of the differential 
evolution equations, see, e.g., \cite{grv90}. 
In fact, several of the NLO evolution 
codes used for parton density analyses in the unpolarized and polarized
cases are set up in Mellin-$n$ space. After evolving from one
scale to another in moment space, the evolved parton 
distributions in Bjorken-$x$ space are recovered by an inverse
Mellin transform, given by 
\beq \label{eq6}
 \Delta f_i(x,\mu) = \frac{1}{2 \pi i} \int_{{\cal C}_n} dn \; 
x^{-n} \Delta f_i^n(\mu) \; , 
\eeq
where ${\cal C}_n$ denotes a contour in the complex $n$ plane that 
has an imaginary part ranging from $-\infty$ to $\infty$ and that 
intersects the real axis to the right of the rightmost poles of 
the $\Delta f_i^n(\mu)$. The evolution of the time-like fragmentation
functions can be treated in a very similar way in Mellin space as well.

The crucial, but simple, step in applying moment techniques to
Eq.~(\ref{eq1}) is to express the $\Delta f_i(x_i,\mu_F)$
by their Mellin inverses in Eq.~(\ref{eq6})
\cite{ref:kosower}. One subsequently interchanges integrations 
and arrives at
\begin{eqnarray} \label{eq7}
\frac{d\Delta \sigma^H}{dO}
&=& 
\frac{1}{(2 \pi i)^2}  \sum_{a,b,c} \, \int_{{\cal C}_n} dn \; 
\int_{{\cal C}_m} dm\;\Delta f_a^n(\mu_F)\,\Delta f_b^m(\mu_F)
\nonumber \\[2mm]
&& 
\times \, \int_{\mathrm{exp-bin}} dT\, 
\int_{x_a^{\mathrm min}}^1 dx_a 
\int_{x_b^{\mathrm min}}^1 dx_b 
\int_{z_c^{\mathrm min}}^1 dz_c\,\,
x_a^{-n} x_b^{-m}\; D_c^H(z_c,\mu_F')
\nonumber \\[2mm]
&& 
\times \,
\frac{d\Delta \hat{\sigma}_{ab}^{c}}{dOdT}
(x_aP_A,\, x_bP_B\, ,P_H/z_c,T,\,\mu_R,\mu_F,\mu_F') \\[2mm]
&\equiv& 
\sum_{a,b} \, \int_{{\cal C}_n} dn \; 
\int_{{\cal C}_m} dm\;\Delta f_a^n(\mu_F)\,\Delta f_b^m(\mu_F)
\; \Delta \tilde{\sigma}_{ab}^{H} (n,m,O,\mu_R,\mu_F)
\; .\label{eq8}
\end{eqnarray}
One can now pre-calculate the quantities 
$\Delta \tilde{\sigma}_{ab}^{H} (n,m,O,\mu_R,\mu_F)$, 
which do not depend at all on the
parton distribution functions, {\em prior} to the fit for a specific
set of the two Mellin variables $n$ and $m$, for each contributing 
subprocess and in each experimental bin. Effectively, one has to 
compute the cross sections with complex ``dummy'' parton distribution 
functions $x_a^{-n} \, x_b^{-m}$. We emphasize that all the tedious 
and time-consuming integrations are already dealt with in the calculation 
of the $\Delta \tilde{\sigma}_{ab}^{H} (n,m,O,\mu_R,\mu_F)$.
We have included the integration over the fragmentation function
$D_c^H$ and the summation over the final state parton $c$ in the 
definition of the pre-calculated quantities $\Delta \tilde{\sigma}_{ab}^{H}$.
This also implies that $\Delta \tilde{\sigma}_{ab}^{H}$ does not
depend anymore on the choice for $\mu_F'$ apart from some
residual dependence which is of higher order in $\alpha_s$.
Usually the fragmentation functions are taken from ``elsewhere'',
i.e., $e^+e^-$ data, rather than being fitted simultaneously
with the parton densities.
We note, however, that one can also replace $D_c^H$ by their
Mellin inverse according to Eq.~(\ref{eq6}). In that case
the pre-calculated quantities would depend on three Mellin
variables.

\begin{figure}[ht] 
\hspace*{-1.5cm}
\epsfig{file=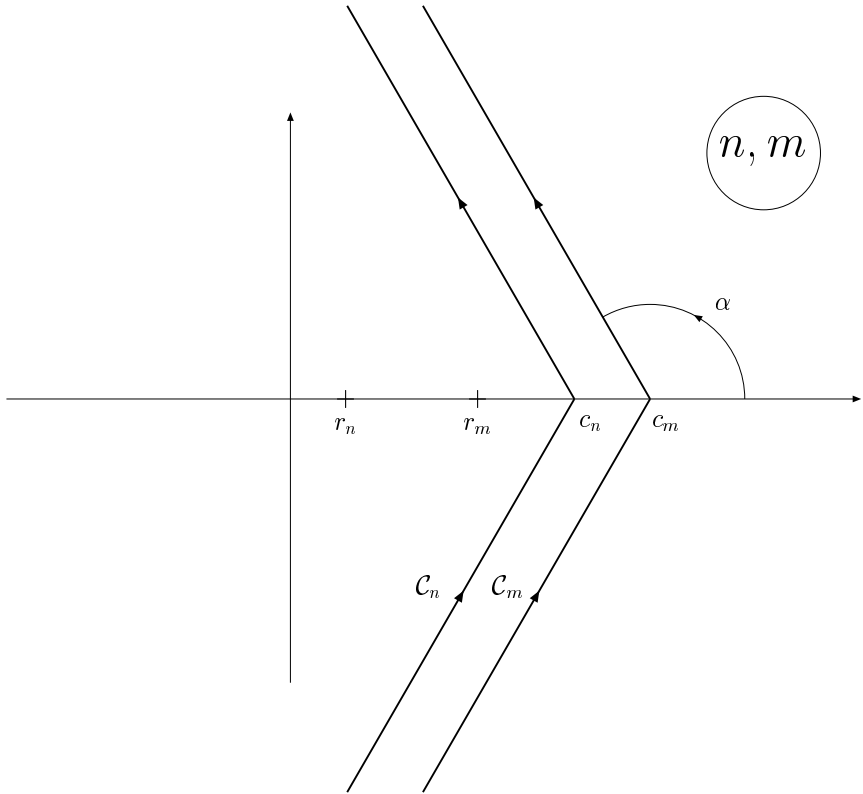,width=15.0cm}
\vspace*{-9.0cm}
\caption{\sf Contours in complex Mellin-$n,m$ spaces for the 
calculation of the double Mellin inverse in Eq.~(\ref{eq8}). 
$r_n$ and $r_m$ denote the rightmost poles of the integrand in $n$ and $m$,
respectively, and the $c_i$ the intersections with the real axis. 
\label{fig1}}
\end{figure}

The double inverse Mellin transformation which finally links the 
parton distributions with the pre-calculated 
$\Delta \tilde{\sigma}_{ab}^{H} (n,m,O,\mu_R,\mu_F)$
of course still needs to be performed 
{\em in each step} of the fitting procedure. However,
the integrations over $n$ and $m$ in Eq.~(\ref{eq8}) 
are extremely fast to perform by  
choosing the values for $n,m$ in  
$\Delta \tilde{\sigma}_{ab}^{H} (n,m,O,\mu_R,\mu_F)$
on the contours ${\cal C}_n$,
${\cal C}_m$ simply as the supports for a Gaussian integration. 
The point here is that the integrand in $n$ and $m$ falls off very 
rapidly as $|n|$ and $|m|$ increase along the contour, for two reasons:
first, each parton distribution function is expected to fall off at
least as a power $(1-x)^3$ at large $x$, which in moment space
converts into a fall-off of $\sim 1/n^4$ or higher. Secondly, 
we may choose contours in moment space that are bent by an 
angle $\alpha-\pi/2$ with respect to the vertical direction;
a possible choice is shown in Fig.~\ref{fig1}. Then, for large
$|n|$ and $|m|$, $n$ and $m$ will acquire large negative real parts, 
so that $(x_a)^{-n}$ and $(x_b)^{-m}$ decrease
exponentially along the respective contours. 
This helps for the numerical convergence of the calculation 
of the $\Delta \tilde{\sigma}_{ab}^{H} (n,m,O,\mu_R,\mu_F)$
and also gives them a rapid fall-off at large 
arguments. We note that no new poles in $n$ and $m$, beyond those
already present in the moments of the parton distribution functions, are 
introduced by the $\Delta \tilde{\sigma}_{ab}^{H} (n,m,O,\mu_R,\mu_F)$
\cite{ref:kosower}.

We note that if one wishes to integrate also over an experimental
bin in $O$ in Eq.~(\ref{eq7}), a potential complication arises
if the hard scale $\mu_F$ in the parton distribution
functions depends explicitly on $O$.
This makes it impossible to straightforwardly
include the $O$ integration
in the pre-calculation of the $\Delta \tilde{\sigma}_{ab}^{H}$. 
A typical example for $O$, which often appears in practice,
is the transverse momentum $p_T$ of an observed 
jet, hadron, or prompt photon.
In this case, the $O$ dependence of $\mu_F$ is,
however, not a serious limitation~\cite{bghv}: 
the logarithmic dependence
of the parton densities on $\mu_F$ is much weaker than the 
overall $p_T$ dependence of the cross section. Therefore, it is 
always possible to choose a bin-average of $p_T$ as the scale
in the parton densities. Alternatively, one could choose not
to include the $p_T$ integration in the $\Delta 
\tilde{\sigma}_{ab}^{H}$ and to construct
grids of somewhat larger size, taken at a small number of support 
points for a simple Gaussian integration over the $p_T$ bin. 
A further possibility \cite{ref:kosower} is to absorb also the evolution of 
the parton densities from their initial scale $\mu_0$ to 
$\mu_F$ into the $\Delta \tilde{\sigma}_{ab}^{H}$, 
which in moment space simply enters in the form of exponentials 
involving the anomalous dimensions, see, e.g., \cite{grv90}. 
This procedure, 
which is somewhat more involved, would eliminate any complication
related to $\mu_F\sim {\cal O}(p_T)$. Anyway, the experiments 
will usually quote results for the $p_T$-{\em differential} cross 
section at the $p_T$-average over the bin, which of course is exactly what 
we have considered in Eqs.~(\ref{eq7}), (\ref{eq8}).
In the latter case it is also easily possible
to organize the grids in such a way that the renormalization and/or 
factorization scales can be varied during the fit, by 
simply taking the (logarithmic) dependence on $\mu_{R,F}$ and the 
strong coupling $\alpha_s(\mu_R)$ out of the partonic 
cross sections beforehand.

As a technical sidestep, we give an explicit expression for
the double inverse transform in Eq.~(\ref{eq8}) for the contours
depicted in Fig.~\ref{fig1}. To this end, we parameterize 
the various segments in Fig.~\ref{fig1} by
\begin{equation}
n = c_n + u_n \,{\rm e}^{\pm i\alpha} \;\;\;{\mathrm and}\;\;\;
m = c_m + u_m \,{\rm e}^{\pm i\alpha} \; ,
\end{equation}
where $u_{n,m}\in [0,\infty]$ and the sign of $\alpha$ has to be
chosen appropriately for the branches of the contours. We then
find
\beqa \label{eq10}
\frac{d\Delta \sigma^H}{dO}
&=&
-\frac{1}{2\pi^2} \sum_{a,b} \,{\rm Re} \Bigg[
\int_0^{\infty} du_n \int_0^{\infty} du_m\;\Delta f_a^n(\mu_F)\, \\[2mm]
&&
\times \Bigg\{ {\rm e}^{2 i\alpha} 
\Delta f_b^m(\mu_F)\,
\Delta \tilde{\sigma}_{ab}^{H} (n,m,O,\mu_R,\mu_F)
-\left(\Delta f_b^{m}(\mu_F)\right)^*\,
\Delta \tilde{\sigma}_{ab}^{H} (n,m^*,O,\mu_R,\mu_F)
\Bigg\}\Bigg] \; \nonumber ,
\eeqa
where the asterisk denotes the complex conjugate, and where we
have made use of $\Delta f_b^{m^*}(\mu_F)=\left(\Delta f_b^{m}(\mu_F)
\right)^*$, since the $\Delta f_i(x,\mu)$ are real functions. 
This identity also
implies that there is no need to separately compute the moments of the 
parton densities at the complex conjugate values $n^*,m^*$, which
has a further positive effect on the computing time required for performing 
the Mellin inverses.
In addition, there is no need to provide separate grids for
$\Delta \tilde{\sigma}_{ab}^{H} (n^*,m,O,\mu_R,\mu_F)$ and
$\Delta \tilde{\sigma}_{ab}^{H} (n^*,m^*,O,\mu_R,\mu_F)$.

Before proceeding, we reemphasize that the idea outlined above
of reverting to Mellin moment space in the 
implementation of any higher-order cross section into parton density 
fits is not entirely new, but was first developed in
Refs.~\cite{bghv,ref:kosower}. 
The example considered in \cite{bghv} was jet production 
in DIS, which offers the simplification 
of being only linear in the parton distribution functions. 
There is a difference between our approach and that of 
Ref.~\cite{bghv} in practical terms: in the language of our 
example in Eq.~(\ref{eq7}), Ref.~\cite{bghv} would insert a factor 
$(x_a^{\rm min})^n\, (x_b^{\rm min})^m$ in the integrands for the 
$x_a$ and $x_b$ integrations, while undoing this operation through a 
factor $(x_a^{\rm min})^{-n}\, (x_b^{\rm min})^{-m}$ in the 
$n$ and $m$ integrands. Even though obviously equivalent mathematically, 
the disadvantage of this procedure is that the resulting factors
$(x_a^{\rm min}/x_a)^n$, $(x_b^{\rm min}/x_b)^m$ in the $x_a$, $x_b$
integrands will now {\em grow}
exponentially along the contours in Mellin space, making it numerically 
much more cumbersome~\cite{bghv} to perform the $x_a$ and $x_b$ integrations 
yielding the $\Delta \tilde{\sigma}_{ab}^{H} (n,m,O,
\mu_R,\mu_F)$. As a matter of fact, the actual extension of the method of
Ref.~\cite{bghv} to the case of hadronic collisions involving bilinear
combinations of parton distributions appears difficult. 
The generalization of \cite{bghv} to hadron-hadron
scattering, without the shortcomings mentioned above,
was first provided in \cite{ref:kosower}.
The difference between our organization of the expression in Eq.~(\ref{eq8})
and Ref.~\cite{ref:kosower} is the choice of the contour.
Reference \cite{ref:kosower} fully exploits the freedom in deforming
the contours for the inverse Mellin transform and constructs 
a ``surface of steepest descent'' which in principle has the
best numerical convergence properties but is difficult to
parameterize.
Instead we stick to the simple contours in Fig.~\ref{fig1} which,
as we will show below, turn out to be sufficient 
to obtain numerical agreement between Eqs.~(\ref{eq1}) and (\ref{eq8}) 
of far better than 1\% for all applications we are going
to consider.
We should also note that in \cite{ref:kosower} the usefulness
of the Mellin transform method was not demonstrated in practice.

%%%%%%%%%%%%%%%%%%%%%%%%%%%%%%%%%%%%%%%%%%%%%%%%%%
\section{Semi-Inclusive Deep-Inelastic Scattering}
%%%%%%%%%%%%%%%%%%%%%%%%%%%%%%%%%%%%%%%%%%%%%%%%%%
%
As a first application for the Mellin transform technique
outlined in the previous Section we consider the semi-inclusive
production of a hadron $H$ in DIS.
SIDIS starts at the Born level with the LO reaction $\gamma^* q\rightarrow q$.
The NLO ${\cal O}(\alpha_s)$ corrections also
comprise the processes $\gamma^* q\rightarrow q g$ and
$\gamma^* g \rightarrow q\bar{q}$ and have been calculated in 
the spin-dependent case in the $\overline{\mathrm{MS}}$ scheme in
\cite{ref:lambda}.
In each case one of the final state partons subsequently fragments into
the observed hadron $H$.
As in the fully inclusive case, the expression for the cross
section is given by a single structure function
$g_{1}^{H}(x,z,Q)$:
\begin{equation}
\label{eq:g1sidis}
\frac{d\Delta\sigma^H}{dx\, dy\, dz} = \frac{4 \pi\alpha^2}{Q^2}
(2-y)\, g_1^{H}(x,z,Q)  \; .
\end{equation}
To NLO in $\alpha_s$, $g_1^{H}$ can be 
written as \cite{ref:lambda,ref:daniel}
\begin{eqnarray}
\label{eq:polsidis}
2\, g_{1}^{H}(x,z,Q) 
&=& 
\sum_{q=u,\bar{u},\ldots,\bar{s}} \!\!\! e_q^2 \;
\Bigg[ \Delta q \left(x,\mu_F\right) D_q^H\left(z,\mu_F'\right) +
\frac{\alpha_s(\mu_R)}{2\pi}
\int_x^1 \frac{d\hat{x}}{\hat{x}} \int_z^1 \frac{d\hat{z}}{\hat{z}}\Bigg\{
\nonumber  \\ [2mm]
\nonumber
&&  
\Delta q \left(\frac{x}{\hat{x}},\mu_F\right)
\Delta C_{qq}^{(1)}(\hat{x},\hat{z},\frac{\mu_F}{Q},\frac{\mu_F'}{Q})
D_{q}^H\left(\frac{z}{\hat{z}},\mu_F'\right)+ 
\nonumber \\[2mm]
&&
\Delta q \left(\frac{x}{\hat{x}},\mu_F\right)
\Delta C_{gq}^{(1)} (\hat{x},\hat{z}, \frac{\mu_F}{Q},\frac{\mu_F'}{Q})
D_{g}^H\left(\frac{z}{\hat{z}},\mu_F'\right) +
\nonumber\\ [2mm]
&&
\Delta {g} \left(\frac{x}{\hat{x}},\mu_F\right)
\Delta C_{qg}^{(1)} (\hat{x},\hat{z},\frac{\mu_F}{Q},\frac{\mu_F'}{Q})
D_{q}^H\left(\frac{z}{\hat{z}},\mu_F'\right)\Bigg\}\Bigg]
\end{eqnarray}
with $x$ and $y$ denoting the usual DIS scaling variables
($Q^2=-q^2=x y S$), and where \cite{ref:aempi,ref:zdef}
$z\equiv p_H\cdot p_N/p_N\cdot q$.
Eq.~(\ref{eq:polsidis}) and the variable $z$ only
apply to hadron production in the current fragmentation region
characterized by positive values for the Feynman variable $x_F$.
All NLO $\overline{\mathrm{MS}}$ partonic coefficient functions 
$\Delta C_{ij}^{(1)}$ are collected in App.~C of \cite{ref:lambda}.
They are non-trivial functions of $x$ {\em and} $z$ such that
the $x$ and $z$ dependences of the cross section do not factorize into
separate functions.
Therefore the inclusion
of the NLO corrections seems to be indispensable for a reliable
extraction of parton densities from SIDIS.

Due to the double convolutions appearing in Eq.~(\ref{eq:polsidis}) and
the fact that the coefficient functions contain mathematical distributions
as $x\rightarrow 1$ and/or $z\rightarrow 1$, the direct use
of Eq.~(\ref{eq:polsidis}) in a global analysis of parton
densities is rather time consuming and awkward, though not
impossible \cite{ref:danielsidis}
since the  partonic coefficient functions are still fairly simple.
SIDIS has the advantage, however, that the Mellin moments in $x$ and $z$
can be taken completely analytically for the partonic
coefficient functions in Eq.~(\ref{eq:polsidis}). In doing so,
the double convolutions in Eq.~(\ref{eq:polsidis}) reduce to
simple multiplications 
$\sim \Delta f_j^n(\mu_F) \Delta C_{ij}^{(1),nm}(\frac{\mu_F}{Q},\frac{\mu_F'}{Q})
D_i^m(\mu_F')$ and all distributions become ordinary functions
of the moment variables.
The $\Delta C_{ij}^{(1),nm}$, defined by
\begin{equation}
\label{eq:cmellin}
\Delta C_{ij}^{(1),nm}(\frac{\mu_F}{Q},\frac{\mu_F'}{Q}) \equiv
\int_0^1 dx\, x^{n-1} \int_0^1 dz\, z^{m-1} 
\Delta C_{ij}^{(1)}(x,z,\frac{\mu_F}{Q},\frac{\mu_F'}{Q})\;\;,
\end{equation}
are straightforwardly determined
from the expressions for the 
corresponding $\Delta C_{ij}^{(1)}(x,z,\frac{\mu_F}{Q},\frac{\mu_F'}{Q})$
in App.~C of \cite{ref:lambda} and read:
\begin{eqnarray} 
\label{cqq}
\Delta C_{qq}^{(1),nm}(\frac{\mu_F}{Q},\frac{\mu_F'}{Q}) 
\!&=&\!
C_F \Bigg[
-8-\frac{1}{m^2} +\frac{2}{(m+1)^2}+\frac{1}{n^2}+
\frac{(1+m+n)^2-1}{m (m+1)n(n+1)}+3 S_2(m)   \nonumber \\ [1.5mm]
&&
 - S_2(n) + \left[ S_1(m) + S_1(n) \right] \left\{ 
S_1(m) + S_1(n) -\frac{1}{m(m+1)}-\frac{1}{n(n+1)}\right\} \nonumber \\ [1.5mm]
&&
+ \left[\frac{2}{n(n+1)}+3-4 S_1(n)\right] \ln\left(\frac{Q}{\mu_F}\right)
 \nonumber \\ [1.5mm] 
&&
+ \left[\frac{2}{m(m+1)}+3-4 S_1(m)\right] \ln\left(\frac{Q}{\mu_F'}\right)
\Bigg] \; ,  \\ [2mm] 
\label{cgq}
\Delta C_{gq}^{(1),nm} (\frac{\mu_F}{Q},\frac{\mu_F'}{Q})
\!&=&\! 
C_F \Bigg[
\frac{2-2 m-9 m^2+m^3-m^4+m^5}{m^2 (m-1)^2(m+1)^2}+
\frac{2m}{n (m+1)(m-1)}  \nonumber \\ [1.5mm]
&& 
-\frac{2-m+m^2}{m (m+1)(m-1)(n+1)}-
\frac{2+m+m^2}{m (m+1)(m-1)} 
\left[ S_1(m) + S_1(n) \right]  \nonumber \\ [1.5mm]
&&
- \frac{2}{(m+1)n (n+1)}
+ 2 \frac{2+m+m^2}{m(m+1)(m-1)} \ln\left(\frac{Q}{\mu_F'}\right)
\Bigg] \; , \\ [2mm] \label{cqg}
\Delta C_{qg}^{(1),nm} (\frac{\mu_F}{Q},\frac{\mu_F'}{Q})
\!&=&\!
T_R \frac{n-1}{n(n+1)}
\Bigg[\frac{1}{m-1}-\frac{1}{m}+\frac{1}{n}-S_1(m) - S_1(n)
+ 2 \ln\left(\frac{Q}{\mu_F}\right) \Bigg] \; ,
\end{eqnarray}
where $C_F=4/3$, $T_R=1/2$, and 
\begin{equation}
S_i(n) \equiv \sum_{j=1}^n \frac{1}{j^i} \; .
\end{equation}

For completeness we give also the Mellin moments for
the corresponding unpolarized coefficient functions 
$C_{1,ij}^{(1),nm}$ and $C_{L,ij}^{(1),nm}$ relevant
for the structure functions $F_1^H$ and $F_L^H$, respectively. 
Using again Eq.~(\ref{eq:cmellin}) and the $x, z$ space 
expressions in App.~C in \cite{ref:lambda} one finds
\begin{eqnarray}
C_{1,qq}^{(1),nm} (\frac{\mu_F}{Q},\frac{\mu_F'}{Q})
&=&
\Delta C_{qq}^{(1),nm}  (\frac{\mu_F}{Q},\frac{\mu_F'}{Q}) +
C_F \frac{2}{m(m+1)n(n+1)}\;\;, \\ [1.5mm]
C_{1,gq}^{(1),nm} (\frac{\mu_F}{Q},\frac{\mu_F'}{Q})
&=&
\Delta C_{gq}^{(1),nm} (\frac{\mu_F}{Q},\frac{\mu_F'}{Q}) +
C_F \frac{2}{(m+1)n(n+1)}\;\;,  \\ [1.5mm]
C_{1,qg}^{(1),nm} (\frac{\mu_F}{Q},\frac{\mu_F'}{Q})
&=&
\nonumber
T_R \Bigg[\frac{2+n+n^2}{n(n+1)(n+2)} \Bigg(
\frac{1}{m-1}-\frac{1}{m}-S_1(m)-S_1(n)+2 \ln\left(\frac{Q}{\mu_F}\right)
\Bigg)  \\ [1.5mm]
&&
\hspace*{6mm}+\frac{1}{n^2} \Bigg]\;\;,
\end{eqnarray}
and
\begin{eqnarray}
C_{L,qq}^{(1),nm} &=& C_F \frac{4}{(m+1)(n+1)}\;\;, \\ [1.5mm]
C_{L,gq}^{(1),nm} &=& C_F \frac{4}{m(m+1)(n+1)}\;\;, \\ [1.5mm]
C_{L,qg}^{(1),nm} &=& T_R \frac{8}{(n+1)(n+2)}\;\;.
\end{eqnarray}
We note that the usefulness of taking double Mellin moments for
unpolarized SIDIS was first pointed out,
though not further pursued, in \cite{ref:aempi}.
In the polarized case Mellin-$n$ moments of the semi-inclusive cross
section at fixed $z$ have been recently considered
in \cite{ref:sisakian}.

Having available the coefficient functions in Mellin moment space
one can evaluate the desired SIDIS structure
function $g_1^H$ in a fast way
by a double inverse Mellin transform as discussed in
Section 3. One further ingredient required is the evolution
of the moments of the fragmentation functions $D_i^m(\mu_F')$
which proceeds along very similar lines as for the parton densities.
Below we will use the recent NLO analysis of
\cite{ref:kretzer} which can be applied down to the $Q$ values required
for the available spin-dependent SIDIS fixed-target data
\cite{ref:smc,ref:hermes}. 
It should be noted that the Mellin
approach allows in principle a simultaneous fit of parton 
densities and fragmentation functions in SIDIS at no
extra ``costs''. 

\begin{figure}[th]
\vspace*{-0.8cm}
\begin{center}
\epsfig{file=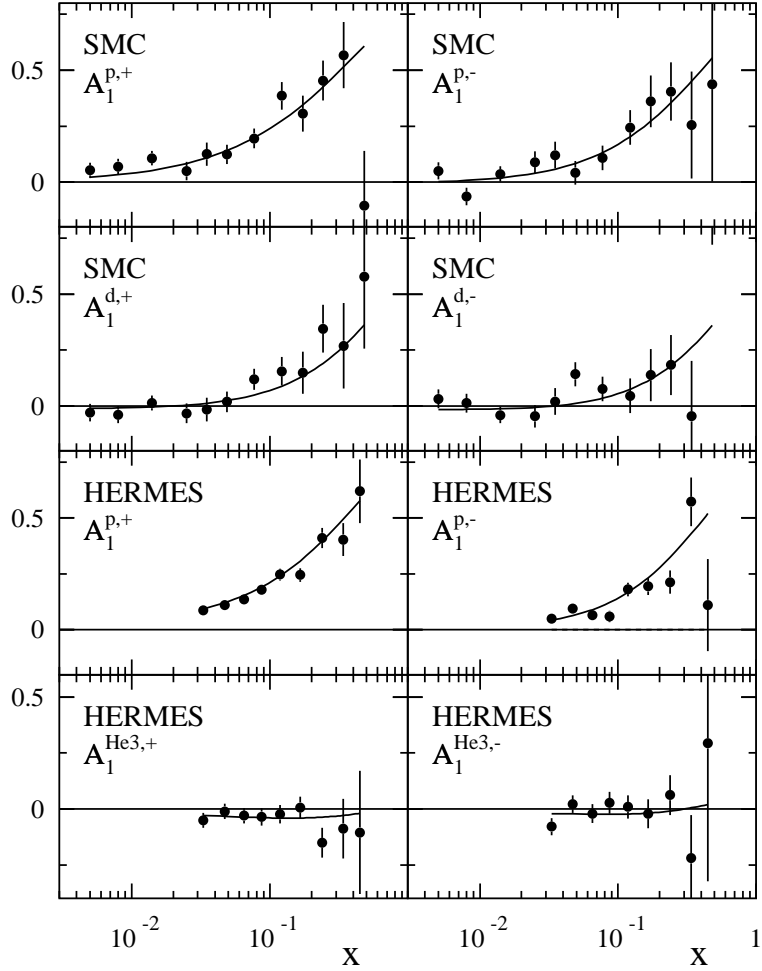,width=0.7\textwidth}
\end{center}
\vspace*{-0.8cm}
\caption{\label{fig:sidis} \sf Comparison of a fit to DIS and
SIDIS data in NLO QCD (see text) with the 
measured SIDIS spin asymmetries $A^{N,\pm}_1$ 
for the production of positively or negatively charged hadrons ($\pm$) off
different targets $N$ \cite{ref:smc,ref:hermes}.}
\end{figure}
The experimentally relevant quantity is the so-called spin
asymmetry, defined as the ratio of the polarized and
unpolarized SIDIS structure functions, $g_1^H(x,z,Q)$ and
$F_1^H(x,z,Q)$, respectively,
\begin{equation}
\label{eq:a1sidis}
A_1^H(x,z,Q) = \frac{g_1^H(x,z,Q)}{F_1^H(x,z,Q)}\;\;.
\end{equation}
\begin{figure}[th]
\begin{center}
\epsfig{file=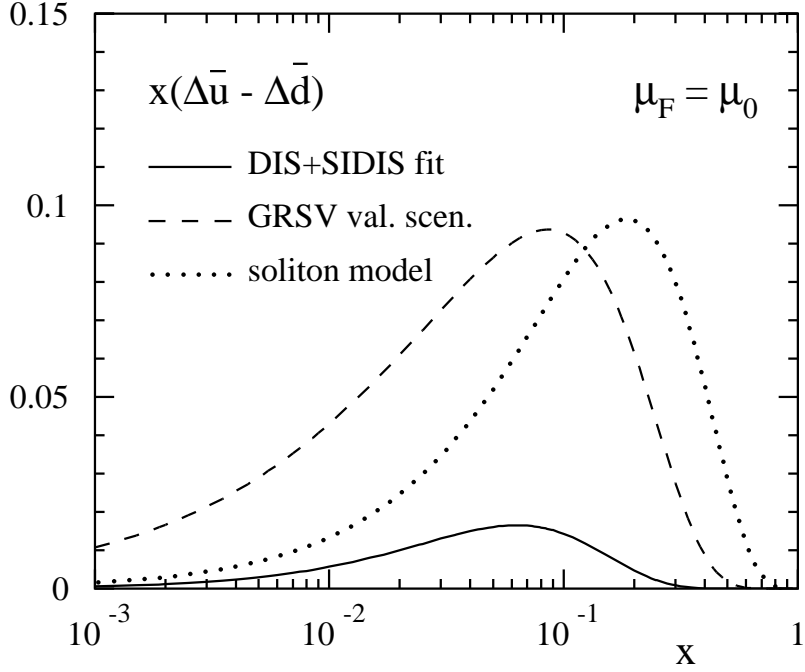,width=0.75\textwidth}
\end{center}
\vspace*{-0.8cm}
\caption{\label{fig:su2} \sf Resulting flavor asymmetry of the
light sea, $\Delta \bar{u} - \Delta \bar{d}$, at the input scale $\mu_0$ from our combined
fit to DIS and SIDIS data. Also shown are model predictions
taken from \cite{ref:su2br1} (dotted line) and 
\cite{ref:su2br3,grsv} (dashed line).}
\end{figure}
Due to limited statistics all presently available results
for Eq.~(\ref{eq:a1sidis}) are integrated over the entire
$z$ range accessible experimentally ($z>0.2$) \cite{ref:smc,ref:hermes}.
To facilitate the comparison with these data it is more
convenient to define an ``effective'' coefficient function
$\Delta \tilde{C}$ rather than using the double moments and integrating
afterwards over $z$. The $\Delta \tilde{C}$ already
incorporate the $z$ integration and can be easily
pre-calculated {\em once} prior to the fit.
They are defined by
\begin{equation}
\Delta \tilde{C}_j^{(1),n}(\frac{\mu_F}{Q}) \equiv
\int_{z_{\mathrm{min}}}^1 dz \int_{{\cal{C}}_m}dm\,
z^{-m}\, \Delta C_{ij}^{(1),nm}(\frac{\mu_F}{Q},\frac{\mu_F'}{Q})\, 
D_i^m(\mu_F')
\end{equation}
(in LO one has $\Delta C_{ij}^{(0),nm}=1$) and can be used
in a similar way as the usual fully inclusive DIS coefficient
functions. This makes the numerical evaluation extremely fast:
100 calculations of the SIDIS cross section in NLO take only
about 1 second on a standard workstation. Clearly, SIDIS data
can be as easily incorporated in a global QCD analysis as
DIS data.

In Fig.~\ref{fig:sidis} we compare the result of a NLO fit
to all available data for DIS {\em and} SIDIS spin asymmetries
with data for $A^{N,\pm}_1$ for positively or negatively charged hadrons 
$H^{\pm}$ and different targets $N$ \cite{ref:smc,ref:hermes}.
Regarding the details of the analysis, we stay in
the framework of the ``standard'' fit of \cite{grsv}, but allow for
an SU(2) breaking of the light sea by introducing a
function $f_{\mathrm{SU(2)}}$, 
\begin{eqnarray}
\Delta u'(x,\mu_0) = \Delta u(x,\mu_0) - 
f_{\mathrm{SU(2)}}(x,\mu_0)\;\;  &,& \;\;
\Delta \bar{u}'(x,\mu_0) = \Delta \bar{u}(x,\mu_0) + 
f_{\mathrm{SU(2)}}(x,\mu_0)\;\;, \nonumber \\ [2mm]
\Delta d'(x,\mu_0) = \Delta d(x,\mu_0) + 
f_{\mathrm{SU(2)}}(x,\mu_0)\;\; &,& \;\;
\Delta \bar{d}'(x,\mu_0) = \Delta \bar{d}(x,\mu_0) - 
f_{\mathrm{SU(2)}}(x,\mu_0)\;\;,
\end{eqnarray}
such that all quark combinations 
measured in inclusive DIS remain unchanged,
$\Delta q' + \Delta \bar{q}' =\Delta q + \Delta \bar{q}$, 
but
$\Delta \bar{u}' - \Delta \bar{d}' = 2f_{\mathrm{SU(2)}}$.
We choose a ``minimal'' ansatz for $f_{\mathrm{SU(2)}}$ 
with 3 additional parameters
\begin{equation}
f_ {\mathrm{SU(2)}}(x,\mu_0) = N x^{\alpha} (1-x)^{\beta}
\end{equation}
where $\mu_0\simeq 0.6\,\mathrm{GeV}$ is the initial scale for the
evolution in \cite{grsv}.
We choose the renormalization and factorization scales
$\mu_R=\mu_F=\mu_F'=Q$.

The resulting asymmetry of the light sea at the input scale
$\mu_0$ is shown in Fig.~\ref{fig:su2}. For comparison we
also show model predictions for $\Delta \bar{u} - \Delta \bar{d}$
from \cite{ref:su2br1} and \cite{ref:su2br3,grsv}.
It turns out that the flavor asymmetry obtained in our analysis
is much less pronounced than predicted in most models.
It has to be stressed, however, that the change in 
the total $\chi^2$ for all SIDIS data is {\em less than one unit}
if one chooses an SU(3) symmetric sea, the model calculations
\cite{ref:su2br1,ref:su2br3,grsv}
or our fit result. Thus one has to conclude that present SIDIS data
are not precise enough to distinguish between different
results for $\Delta \bar{u} - \Delta \bar{d}$ and that one has
to wait for new SIDIS data from HERMES and, in particular, for
results on $W^{\pm}$ boson production at RHIC \cite{rhic}.
Similar conclusions have been reached in 
the analysis of \cite{ref:danielsidis}.

%%%%%%%%%%%%%%%%%%%%%%%%%%%%%%%%%%%%%%%%%%
\section{Prompt Photon Production at RHIC}
%%%%%%%%%%%%%%%%%%%%%%%%%%%%%%%%%%%%%%%%%%
%
To give an example for the Mellin technique in hadron-hadron
collisions, we study the production
of a prompt photon in $pp$ collisions at RHIC. 
In this case $d\Delta \hat{\sigma}_{ab}^{\gamma,(0)}$
in Eq.~(\ref{eq4}) starts at LO with the reactions 
$q+\bar{q}\to \gamma +g$ and $q+g\to \gamma +q$, the latter channel 
being sensitive to the polarized gluon distribution.
The NLO corrections, $d\Delta \hat{\sigma}_{ab}^{\gamma,(1)}$,
are also available~\cite{dgnlo}. The NLO $x$-space expressions are
rather lengthy and complicated, and Mellin moments cannot be taken 
analytically anymore. Nevertheless, as we shall see below, it is in
the analysis of hadron-hadron collision data where the Mellin
moment technique exhibits its full potential and usefulness.

\begin{figure}[ht] 
\hspace*{1.5cm}
\epsfig{file=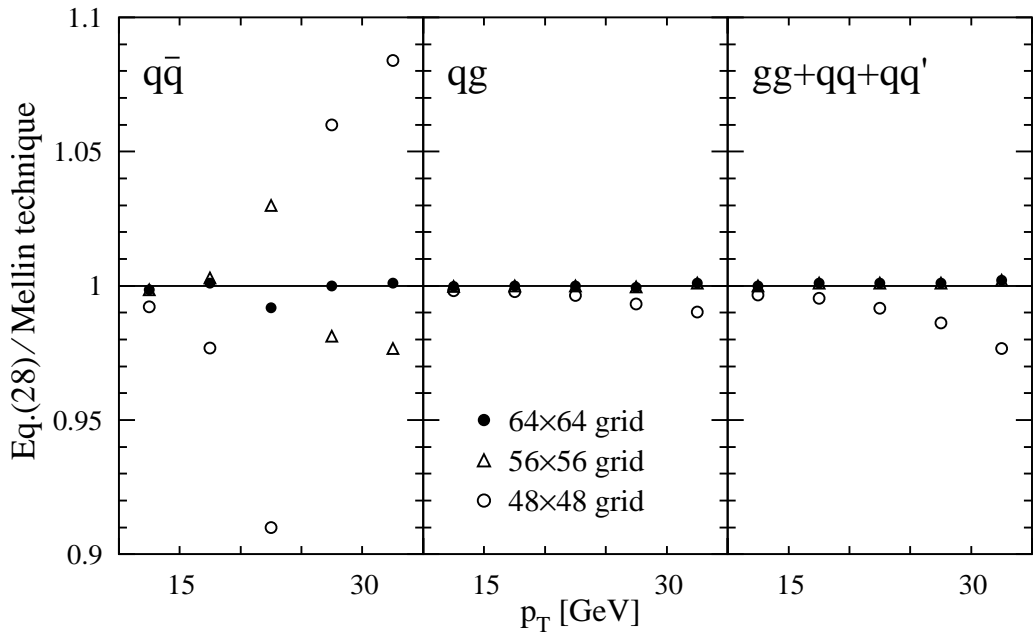,width=15.0cm}
\caption{\sf Comparison of the results based on the Mellin-technique in
Eqs.~(\ref{eq7}) and (\ref{eq8}) to those of Eq.~(\ref{eq:pp}) 
for various sizes of the grid in $n,m$.
\label{fig2}}
\end{figure}
To be specific, the transverse momentum ($p_T$) distribution
of a prompt photon in $pp$ collisions at a center-of-mass energy $\sqrt{S}$, integrated 
over a certain experimental bin in pseudorapidity $\eta$
[i.e., ``$O\equiv p_T$'' and ``$T\equiv \eta$'' in Eq.~(\ref{eq1})],
is given by

\begin{eqnarray} 
\label{eq:pp}
\frac{d\Delta \sigma^{\gamma}}{dp_T} 
&=&
\sum_{a,b}\, \int_{\eta-{\mathrm bin}} d\eta\, 
\int_{x_a^{\mathrm min}}^1 dx_a 
\int_{x_b^{\mathrm min}}^1 dx_b \,\,
\Delta f_a (x_a,\mu_F) \,\Delta f_b (x_b,\mu_F) 
\nonumber \\ [2mm]
&&
\times \,\frac{d\Delta \hat{\sigma}_{ab}^{\gamma}}{dp_T d\eta}
(x_aP_A,\, x_bP_B\, ,p_T,\eta,\,\mu_R,\mu_F) \; ,
\end{eqnarray}
where $x_a^{\mathrm min} = x_T \, {\mathrm e}^{\eta}/(2-x_T \, {\mathrm e}^{-\eta})$
and $x_b^{\mathrm min} = x_a \, x_T \, {\mathrm e}^{-\eta}/(2 x_a-x_T \, 
{\mathrm e}^{\eta})$ with $x_T = 2 p_T/\sqrt{S}$. 
For our case study, we analyse the polarized prompt photon cross section
in Eq.~(\ref{eq:pp}) at NLO. The associated spin-asymmetry, defined
as the ratio of the polarized and the unpolarized cross sections,
\begin{equation}
\label{eq11}
A_{LL}^{\gamma}\equiv \frac{d\Delta \sigma^{\gamma}/dp_T}
{d\sigma^{\gamma}/dp_T} \; ,
\end{equation}
will soon be measured at RHIC in collisions of longitudinally polarized
protons and, as mentioned above, will be a key process for measuring
$\Delta g$. We use $\sqrt{S}=200$ GeV and look at the cross section
as a function of the photon's transverse momentum $p_T$ for five 
values of $p_T$ which will be experimentally accessible at RHIC, 
$p_T=[12.5, 17.5, 22.5, 27.5, 32.5]$ GeV. We average 
over $|\eta|<0.35$ in pseudorapidity. As in experiment~\cite{rhic}, 
we impose an isolation cut on the photon, for which we choose the isolation
proposed in~\cite{frixione} with parameters $R=0.4$, $\epsilon=1$. 
A positive feature of this isolation criterion is the 
absence of a fragmentation contribution to prompt photon production,
hence we can drop the $z_c$ integration and the fragmentation
function $D_c^{\gamma}$ in Eq.~(\ref{eq1}).
We choose the renormalization and factorization scales $\mu_R=\mu_F= p_T$. 

Our first goal here is to show that the method based on Eqs.~(\ref{eq7}) 
and (\ref{eq8}) actually works also for the more complicated
case of hadron-hadron collisions and correctly reproduces the 
result obtained within the direct, but ``slow'', calculation via 
Eq.~(\ref{eq:pp}). Also, we need to establish an optimal size of the grids 
that yields excellent accuracy but is still calculable in, say, a few 
hours of CPU time on a standard workstation. Fig.~\ref{fig2} 
compares the results based on Eqs.~(\ref{eq7}) and (\ref{eq8}), 
referred to as ``Mellin technique'', to those of Eq.~(\ref{eq:pp}), 
for various sizes of the grid in $n,m$. Here we have used again the 
polarized parton densities of~\cite{grsv} (``standard'' set).
For a more detailed comparison, we split up
the contributions to the NLO prompt photon cross section into three
parts, associated with the reactions $q+\bar{q}\to \gamma +X$ and $q+g\to
\gamma +X$ that are already present at Born level, and all other
processes that arise only at NLO. One notices that in each case already 
a grid size of $64\times 64$ values yields excellent accuracy. Even a
$56\times 56$ grid is acceptable apart from a minor deviation 
occuring for $q\bar{q}$ scattering in the vicinty of a zero in the
partonic cross section. We have checked that the results in 
Fig.~\ref{fig2} do not depend on the actual choice of parton densities.

The crucial asset of the Mellin method is the speed at which one can calculate 
the full hadronic cross section, once the grids 
$\Delta \tilde{\sigma}_{ab}^{\gamma} (n,m,p_T,\mu_R,\mu_F)$
have been pre-calculated. For the $64 \times 64$ grid, we found that
1000 evaluations of the full NLO prompt photon cross section
take only about 10-15 seconds on a standard workstation. Note that 
this number includes the evolution (in moment space) of the 
parton distributions from their input scale to the scale $p_T$ 
relevant to this case. Clearly, an implementation into a 
full parton density fitting procedure is now readily possible. 

\begin{figure}[ht]
\begin{minipage}{8.45cm}
\epsfig{figure=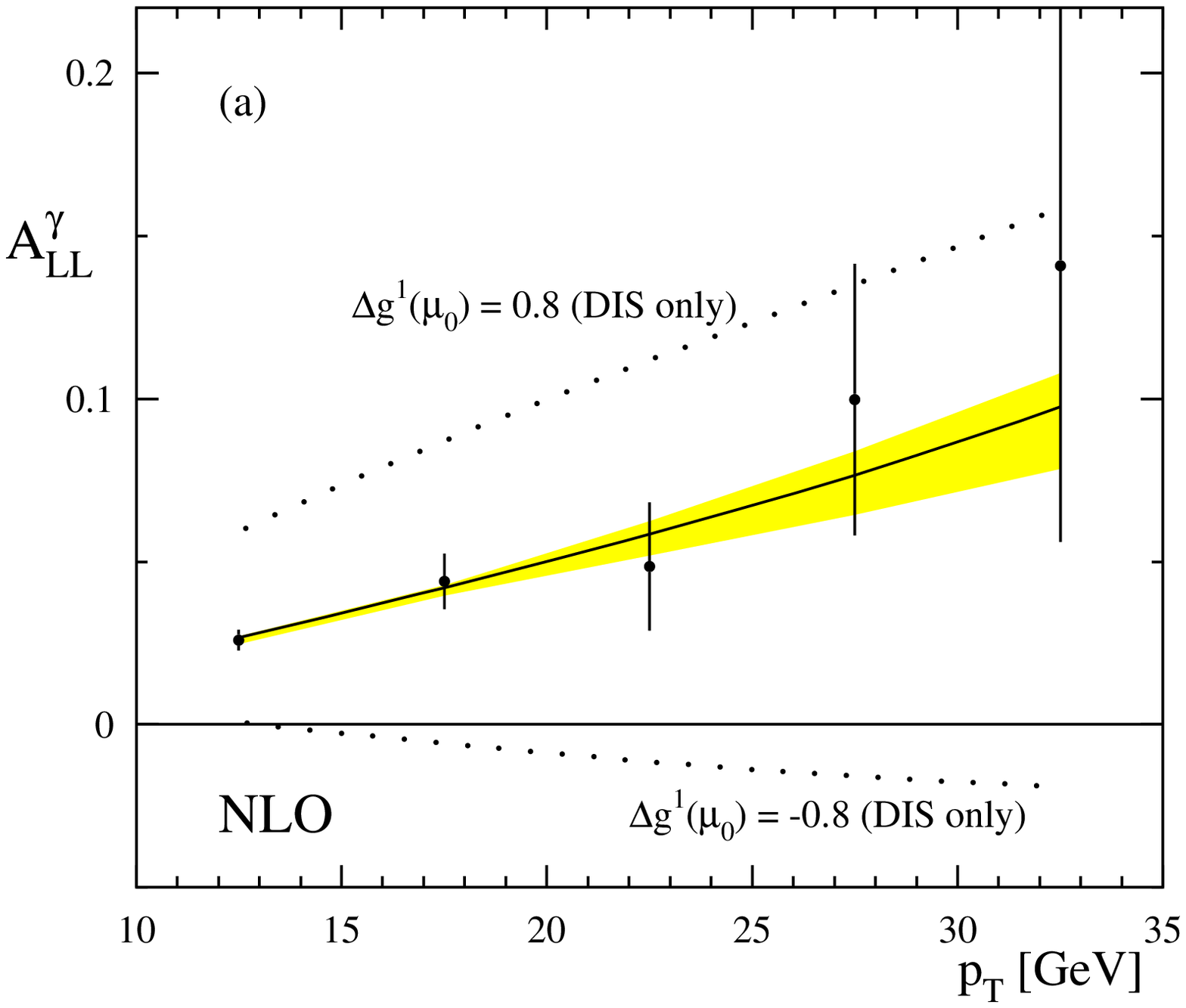,width=1.05\textwidth}
\end{minipage}
\begin{minipage}{8.45cm}
\epsfig{figure=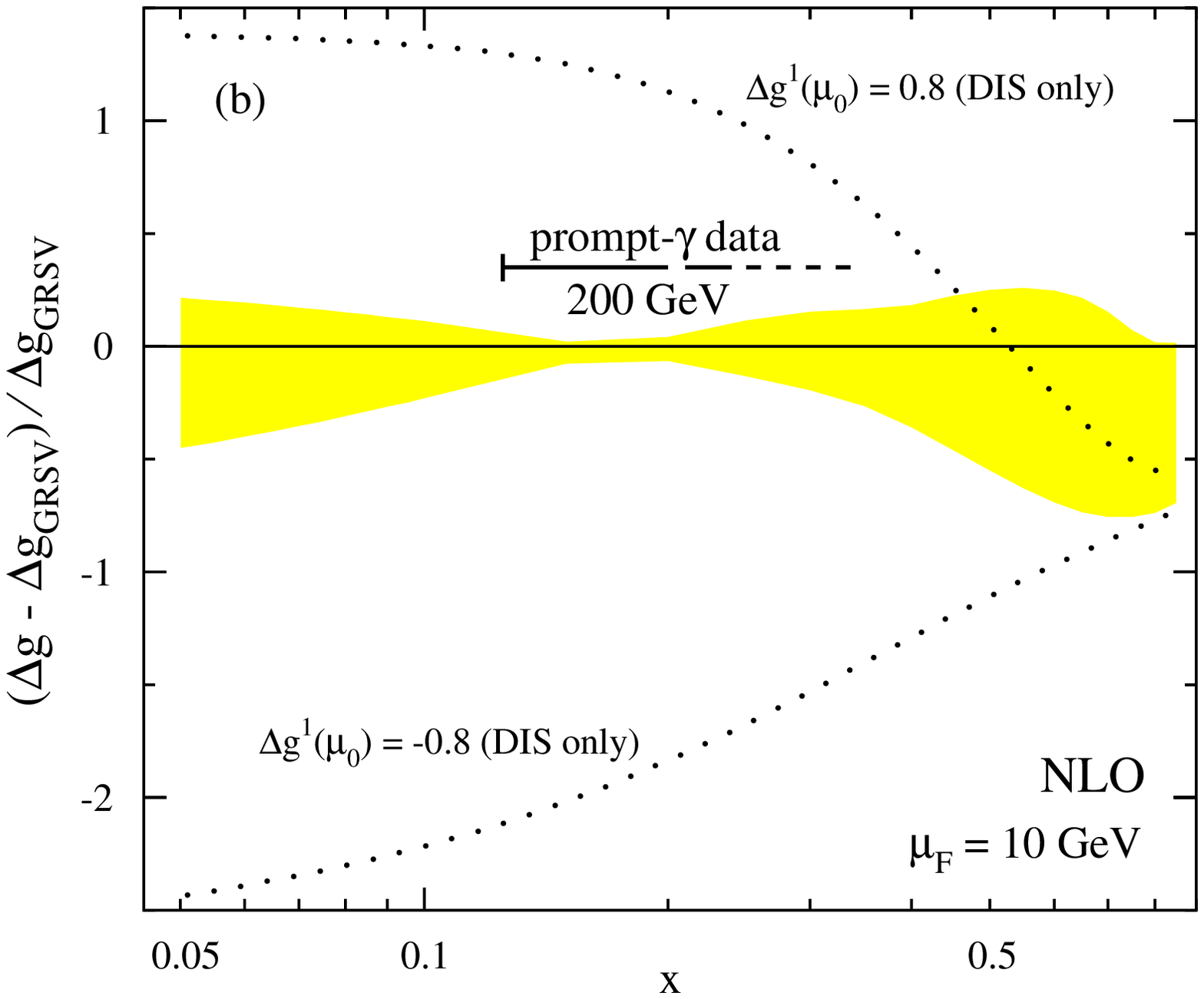,width=1.05\textwidth}
\end{minipage}
\caption{\sf {\bf{(a):}} Generated pseudo data for $A_{LL}^{\gamma}$
based on a calculation using the spin-dependent parton densities of \cite{grsv}
(solid line). The shaded band corresponds to the results of a large number of 
combined fits to DIS and $A_{LL}^{\gamma}$ data (see text).
{\bf{(b):}} Variations (shaded area) of the polarized gluon densities obtained in
the combined fits, with respect to $\Delta g$ of \cite{grsv}, for
$\mu_F=10\,\mathrm{GeV}$ (see text). 
Also shown are two extreme gluon densities (dotted lines) which give an
excellent description of the polarized DIS data only. 
\label{fig3}}
\end{figure}

To give an example, we finally perform a ``toy''
global analysis of the available data on polarized 
DIS~\cite{emc} and of {\em fictitious} data on prompt photon
production at RHIC~\cite{rhic}, which we project by simply calculating 
$A_{LL}^{\gamma}$ in Eq.~(\ref{eq11}) to NLO 
using the sets of polarized and unpolarized parton distributions of \cite{grsv} 
and \cite{grv}, respectively. For an estimate of the anticipated 
$1\sigma$ errors on the ``data'' for $A_{LL}^{\gamma}$, we use the 
numbers reported in~\cite{rhic}. We subsequently apply a random Gaussian 
shift of the pseudo-data, allowing them to vary within $1\sigma$. 
The ``data'', as well as the underlying
theoretical calculation of $A_{LL}^{\gamma}$ based
on the spin-dependent parton densities of \cite{grsv} (solid line), 
are shown in the left panel of Fig.~\ref{fig3}. 

Next, we perform a large number of fits to the full,
{\em DIS plus projected prompt photon}, data set. We simultaneously
fit {\em all} polarized parton densities, (anti)quarks and gluons,
choosing the distributions of \cite{grsv} as the input for the
$\Delta q$, $\Delta \bar{q}$, but using randomly chosen values for 
the parameters in the ansatz for the polarized gluon distribution at 
the input scale $\mu_0$. Regarding the details of the evolution, we stay 
again within the setup of~\cite{grsv}, but we
choose a more flexible ansatz for the polarized gluon density,
\begin{equation} \label{eq12}
\Delta g (x,\mu_0) = N \, x^{\alpha} \, (1-x)^{\beta} \,
(1+\gamma \, x)\; g(x,\mu_0) \; , 
\end{equation} 
which also allows for a zero in the $x$-shape of $\Delta g$.
$g(x,\mu_0)$ is the unpolarized gluon density~\cite{grv} at the
input scale of \cite{grsv}. Note that the functional form 
for the polarized gluon density of~\cite{grsv}, used for generating
our pseudo-data, is included in Eq.~(\ref{eq12}) for $N=1.419,\, 
\alpha=1.43, \, \beta = 0.15,\, \gamma=0$. Each fit takes only about
$10\div 20$ minutes.

Ideally, thanks to the strong sensitivity of the 
prompt photon reaction to $\Delta g$, 
the gluon density in each fit should return close to the function we assumed 
when calculating the fictitious prompt photon ``data'', in the region 
of $x$ probed by the data. Indeed, as shown in Fig.~\ref{fig3}(b), 
this happens. The shaded band illustrates the deviations
of the gluon densities obtained from the global fits to the 
``reference $\Delta g$'' \cite{grsv} used in generating the pseudo-data.
It should be stressed that only those fits are admitted
to the band that give a good simultaneous description of the DIS
{\em and} $A_{LL}^{\gamma}$ data. Here we have tolerated a maximum
increase of the total $\chi^2$ by up to four units from its minimum value.
The shaded area in Fig.~\ref{fig3}(a) shows the corresponding variations in
$A_{LL}^{\gamma}$.

As is expected, all gluon densities are rather tightly constrained in
the $x$-region dominantly probed by the prompt photon data. This is true 
in particular at $x\approx 0.15$, as a result of the most precise data 
point for $A_{LL}^{\gamma}$ at $p_T=12.5\,\mathrm{GeV}$.
We note that one can also easily include the SIDIS data discussed
in Section 3 into the global analysis without any significant increase
of computing time for each fit. However, so far these data have
no impact on our results.
To illustrate our {\em present} ignorance of $\Delta g$, Fig.~\ref{fig3}(b) 
shows also two extreme gluon densities with first moments 
$\Delta g^1(\mu_0)=\pm 0.8$ (dotted lines), which are both in 
perfect agreement with all presently available DIS data. 
The corresponding predictions for 
$A_{LL}^{\gamma}$ for these two sets are given in Fig.~\ref{fig3}(a).
It should be noted that future measurements of $A_{LL}^{\gamma}$ at RHIC
at $\sqrt{S}=500\,\mathrm{GeV}$ and for similar $p_T$ values of the
prompt photon would further reduce the uncertainties on $\Delta g$ 
in the $x$-region between 0.05 and 0.1.
Although our analysis still contains a certain bias by choosing only 
the framework of \cite{grsv} for the fits as well as by our choice of
what $\chi^2$ values are still tolerable, it clearly outlines the 
potential and importance of upcoming measurements of 
$A_{LL}^{\gamma}$ at RHIC for improving
our understanding of the spin structure of the nucleon, in particular
of its spin-dependent gluon density.

%%%%%%%%%%%%%%%%%%%%%
\section{Conclusions}
%%%%%%%%%%%%%%%%%%%%%
%
To conclude, we have presented and applied a powerful technique
for implementing in a fast way, and without any approximations, 
higher-order calculations of partonic cross sections into global 
analyses of parton distribution functions. 
We have demonstrated that the approach works in practice for
two examples: SIDIS and prompt photon production in $pp$ collisions.
In the first case it was possible to perform the Mellin transform
analytically and we have provided all necessary technical details
for future analyses of polarized and unpolarized SIDIS data.
For polarized prompt photon production we have presented a case
study for a future global analysis based on fictitious data.
The Mellin transform method is certainly applicable to any other 
reaction of interest, and it could equally well be an improvement 
also in any global analysis of  unpolarized parton distributions. 

%%%%%%%%%%%%%%%%%%%%%%%%%%
\section*{Acknowledgments}
%%%%%%%%%%%%%%%%%%%%%%%%%%
We are grateful to G.\ Sterman and A.\ Vogt for useful comments, and to
A.\ Deshpande for helpful discussions.  
We also thank S.\ Kretzer for providing us with the
evolution code for the set of fragmentation functions
in \cite{ref:kretzer}.
The work of M.S.\ was supported in part by the National
Science Foundation grant no.\ PHY-9722101.     
W.V.\ is grateful to RIKEN, Brookhaven National Laboratory and the U.S.
Department of Energy (contract number DE-AC02-98CH10886) for
providing the facilities essential for the completion of this work.
%%%%%%%%%%%%%%%%%%%%%%%%%%%

%

\begin{thebibliography}{99}
%%%%%%%%%%%%%%%%%%%%%%%%%%%
%
\bibitem{emc} European Muon Collaboration (EMC), J. Ashman {\it et al.},
                 Phys. Lett. {\bf B206}, 364 (1988);
                 Nucl. Phys. {\bf B328}, 1 (1989);\\
for a recent review of the data on polarized deep-inelastic 
scattering, see: E.\ Hughes and R.\ Voss, 
Annu.\ Rev.\ Nucl.\ Part.\ Sci.\ {\bf 49}, 303 (1999).
%
\bibitem{ref:su2br1} D.\ Diakonov {\em et al.}, Nucl. Phys.
{\bf B480}, 341 (1996); Phys. Rev. {\bf D56}, 4096 (1997);\\
M.\ Wakamatsu and T.\ Kubota, Phys. Rev. {\bf D60}, 034020 (1999);\\
K.\ Goeke {\em et al.}, Acta Phys. Polon. {\bf B32}, 1201 (2001).
%
\bibitem{ref:su2br2} R.S.\ Bhalerao, Phys. Lett. {\bf B380}, 1 (1996);
{\bf B387}, 881 (1996) (E); Nucl. Phys. {\bf A680}, 62 (2000);
Phys. Rev. {\bf C63}, 025208 (2001);\\
R.S.\ Bhalerao, N.G.\ Kelkar, and B.\ Ram, Phys. Lett.
{\bf B476}, 285 (2000);\\
R.J.\ Fries and A.\ Sch\"{a}fer,  Phys. Lett. {\bf B443}, 40 (1998);\\
F.-G.\ Cao and A.I.\ Signal, Eur. Phys. J. {\bf C21}, 105 (2001).
%
\bibitem{ref:su2br3} M.\ Gl\"{u}ck and E.\ Reya, Mod. Phys. Lett.
{\bf A15}, 883 (2000).
%
\bibitem{rhic} for a recent review of the RHIC spin physics program,
see: G.\ Bunce, N.\ Saito, J.\ Soffer, and W.\ Vogelsang,
Annu.\ Rev.\ Nucl.\ Part.\ Sci.\ {\bf 50}, 525 (2000).
%
\bibitem{cteq} CTEQ Collaboration, H.L. Lai {\it et al.},
Eur.\ Phys.\ J.\ {\bf C12}, 375 (2000); \\
A.D.~Martin, R.G.~Roberts, W.J.~Stirling, and R.S.~Thorne, 
Eur.\ Phys.\ J.\ {\bf C4}, 463 (1998).
%
\bibitem{grv}
M.\ Gl\"{u}ck, E.\ Reya, and A.\ Vogt, Eur.\ Phys.\ J.\ {\bf C5}, 461 (1998).
%
\bibitem{bghv} C.\ Berger, D.\ Graudenz, M.\ Hampel, and A.\ Vogt,
Z. Phys. {\bf C70}, 77 (1996).
%
\bibitem{ref:kosower} D.A.\ Kosower, Nucl. Phys. {\bf B520}, 263 (1998).
%
\bibitem{ref:smc} Spin Muon Collaboration (SMC), 
B.\ Adeva {\em et al.}, Phys. Lett. {\bf B420}, 180 (1998). 
%
\bibitem{ref:hermes} HERMES Collaboration, K\ Ackerstaff {\em et al.},
Phys. Lett. {\bf B464}, 123 (1999).
%
\bibitem{fact} S.B.\ Libby and G.\ Sterman, Phys.\ Rev.\ {\bf D18}, 3252 
(1978); \\
R.K.\ Ellis, H.\ Georgi, M.\ Machacek, H.D.\ Politzer, 
and G.G.\ Ross, Phys. Lett. {\bf 78B}, 281 (1978); Nucl. Phys.
{\bf B152}, 285 (1979); \\
D.\ Amati, R.\ Petronzio, and G.\ Veneziano,
Nucl. Phys. {\bf B140}, 54 (1980); Nucl. Phys. {\bf B146}, 29 (1978);\\
G.\ Curci, W.\ Furmanski, and R.\ Petronzio, Nucl.\ Phys.\ {\bf B175}, 27 (1980);\\
J.C.\ Collins, D.E.\ Soper, and G.\ Sterman, Phys.\ Lett.\ {\bf B134}, 263 (1984); 
Nucl.\ Phys.\ {\bf B261}, 104 (1985);\\
J.C.\ Collins, Nucl.\ Phys.\ {\bf B394}, 169 (1993).
%
\bibitem{dglap} G.\ Altarelli and G.\ Parisi, Nucl. Phys. {\bf B126}, 298 
(1977); \\
Yu.L.\ Dokshitser, Sov. Phys. JETP {\bf 46}, 641 (1977); \\ 
L.N.\ Lipatov, Sov. J. Nucl. Phys. {\bf 20}, 95 (1975); \\
V.N.\ Gribov and L.N.\ Lipatov, Sov. J. Nucl. Phys. {\bf 15}, 438 (1972).
%
\bibitem{grv90} M.\ Gl\"{u}ck, E.\ Reya, and A.\ Vogt, Z. Phys. {\bf C48}, 471
(1990). 
%
\bibitem{ref:lambda} D.\ de Florian, M.\ Stratmann, and 
W.\ Vogelsang, Phys. Rev. {\bf D57}, 5811 (1998).
%
\bibitem{ref:daniel} D.\ de Florian, C.A.\ Garcia Canal,
and  R.\ Sassot, Nucl. Phys. {\bf B470}, 195 (1996);\\ 
D. de Florian {\em et al.}, Phys. Lett. {\bf B389}, 358 (1996).
%
\bibitem{ref:aempi} G.\ Altarelli, R.K.\ Ellis, G.\ Martinelli, 
and S.-Y.\ Pi, Nucl. Phys. {\bf B160}, 301 (1979).
%
\bibitem{ref:zdef} W.\ Furmanski and R.\ Petronzio,
Z. Phys. {\bf C11}, 293 (1982);\\
P.\ Nason and B.\ Webber, Nucl. Phys. {\bf B421},
473 (1994); {\bf B480}, 755(E) (1996).
%
\bibitem{ref:danielsidis} D.\ de Florian and R.\ Sassot,
Phys. Rev. {\bf D62}, 094025 (2000).
%
\bibitem{ref:sisakian} A.N.\ Sissakian, O.Yu.\ Shevchenko, 
and V.N.\ Samoilov, {\tt hep-ph/0010298}.
%
\bibitem{ref:kretzer} S.\ Kretzer, 
Phys. Rev. {\bf D62}, 054001 (2000). 
%
\bibitem{grsv}  M.\ Gl\"{u}ck, E.\ Reya, M. Stratmann, and W.\ Vogelsang,
Phys. Rev. {\bf D63}, 094005 (2001). 
%
\bibitem{dgnlo} A.P.\ Contogouris, B.\ Kamal, Z.\ Merebashvili, and 
F.V.\ Tkachov, Phys. Lett. {\bf B304}, 329 (1993); 
Phys. Rev. {\bf D48}, 4092 (1993);\\
A.P.~Contogouris and Z.~Merebashvili, Phys. Rev. {\bf D55}, 2718 (1997);\\
L.E.~Gordon and W.~Vogelsang, Phys. Rev. {\bf D48}, 3136 (1993); 
Phys. Rev. {\bf D49}, 170 (1994); \\
S.~Frixione and W.\ Vogelsang,  
Nucl. Phys. {\bf B568}, 60 (2000).
%
\bibitem{frixione} S.~Frixione, Phys. Lett. {\bf B429}, 369 (1998).
%
\end{thebibliography}
\end{document}